\documentclass[aip,apl,amsmath,amssymb,reprint]{revtex4-1}
                  \usepackage[dvipsnames]{xcolor}

\usepackage[utf8]{inputenc}
\usepackage[T1]{fontenc}
\usepackage{graphicx}
\usepackage{longtable}
\usepackage{wrapfig}
\usepackage{rotating}
\usepackage[normalem]{ulem}
\usepackage{amsmath}
\usepackage{amssymb}
\usepackage{capt-of}
\usepackage{hyperref}

\begin{document}

\title{Broadband polarization insensitivity and high detection efficiency in high-fill-factor superconducting microwire single-photon detectors}
\author{Dileep V. Reddy}
\email{dileep.reddy@nist.gov}
\affiliation{Department of Physics, University of Colorado, Boulder, Colorado 80309, USA}
\affiliation{National Institute of Standards and Technology, Boulder, Colorado 80305, USA}
\author{Negar Otrooshi}
\affiliation{Department of Physics, University of Colorado, Boulder, Colorado 80309, USA}
\affiliation{National Institute of Standards and Technology, Boulder, Colorado 80305, USA}
\author{Sae Woo Nam}
\affiliation{National Institute of Standards and Technology, Boulder, Colorado 80305, USA}
\author{Richard P. Mirin}
\affiliation{National Institute of Standards and Technology, Boulder, Colorado 80305, USA}
\author{Varun B. Verma}
\affiliation{National Institute of Standards and Technology, Boulder, Colorado 80305, USA}

\begin{abstract}
Single-photon detection via absorption in current-biased nanoscale
superconducting structures has become a preferred technology in quantum
optics and related fields. Single-mode fiber packaged devices have seen new
records set in detection efficiency, timing jitter, recovery times, and
largest sustainable count rates. The popular approaches to decreasing
polarization sensitivity have thus far been limited to introduction of
geometrically symmetric nanowire meanders, such as spirals and fractals, in
the active area. The constraints on bending radii, and by extension, fill
factors, in such designs limits their maximum efficiency. The discovery of
single-photon sensitivity in micrometer-scale superconducting wires enables
novel meander patterns with no effective upper limit on fill factor. This
work demonstrates simultaneous low-polarization sensitivity (\(1.02\pm
0.008\)) and high detection efficiency (\(> 91.8\%\) with \(67\%\) confidence at
\(2\times10^5\) counts per second) across a \(40\) nm bandwidth centered at
1550 nm in 0.51 \(\mu\text{m}\) wide microwire devices made of silicon-rich
tungsten silicide, with a \(0.91\) fill factor in the active area. These
devices boasted efficiencies of 96.5-96.9\% $\pm$ 0.5\% at \(1\times10^5\)
counts per second for 1550 nm light.
\end{abstract}
\maketitle

\section{Introduction}
\label{sec:orgd2b1e87}

Superconducting nanowire single-photon detectors (SNSPDs) are a premier
technology for applications that require fast, high-efficient detection and
high-timing resolution. Their utility spans such diverse areas as
fundamental research \cite{Hochberg_2019}, communications
\cite{Mao_2018,Chen_2020}, metrology \cite{Slussarenko_2017}, remote sensing
\cite{Zhu_2017}, materials research \cite{Chen_2018}, and astronomy
\cite{Zhuang_2018,Khabiboulline_2019}. Such detectors in single-mode
fiber-packaged form have been fruitfully employed in several
ground-breaking quantum-optics experiments
\cite{Shibata_2014,Dixon_2014,Shalm_2015,Takesue_2015,Najafi_2015,Jin_2015,Weston_2016,Saglamyurek_2016}.
Within the past five years, fiber-packaged SNSPDs have seen new records set
in such diverse figures-of-merit as system-detection efficiency (SDE)
\cite{reddy2020,Hu2020,Chang2021}, timing jitter \cite{Korzh_2020}, and low
dark counts \cite{Shibata_2017}. The field is making advances towards joint
high performance in multiple metrics simultaneously. One such goal is high
SDE coupled with low polarization sensitivity.

We define polarization sensitivity (PS) for a device as the ratio of the
maximum to minimum SDE across all input polarization states of photons.
Traditional fiber-coupled SNSPDs have consisted of nanowire meanders
covering the active area (where photons are expected to be absorbed) in a
zig-zag pattern. The geometry forms a grating-like structure of parallel
strips of superconductor spaced by a dielectric. Consequently, SNSPDs have
inherently possessed a non-unity PS
\cite{Anant_2008,Redaelli_2016,Zheng_2016,Redaelli_2017}. While such
meanders allow for unity PS at a specific wavelength via cleverly
engineered anti-reflection coatings \cite{reddy2019}, reliable unity-PS
across significant bandwidths has remained unrealized in high-efficiency
devices. Applications that require high efficiency and throughput
\cite{Shalm_2015,Ghafari2019,Shalm2021} often use polarization controllers
before directing light to the detectors, which is a significant source of
loss. High-SDE devices with either unity, or infinite PS (meaning no
sensitivity to one polarization) would mitigate such issues. Such detectors
would also close a security loophole in standard phase-encoding quantum-key
distribution implementations \cite{Wei2019}.

Historical approaches to achieving unity PS have sought to spatially
symmetrize the nanowire-meander geometries. PS values of 1.02-1.04 have
been measured in spiral SNSPDs since 2008
\cite{Dorenbos_2008,Henrich_2013,Huang_2017} with limited SDE. In 2012, Verma
et al. fabricated a two-layer 3D-SNSPD with perpendicularly oriented
meanders, and showed a PS of 1.02 with an SDE of 87.7\% \cite{Verma_2012}.
Space-filling fractals such as Sierpinski or Hilbert curves have also been
studied as a means of introducing discrete rotational symmetries into
nanowire meanders \cite{Gu_2015}. The fractal approach has seen steady
improvement \cite{Chi_2018,Meng_2020}, and has recently realized a PS of
\(1.00\) at 91\% efficiency \cite{Meng20}. The introduction of turns and hairpin
bends in the active area renders the outer-radii regions of such
fractal-meander nanowires relatively insensitive to photons
\cite{Clem_2011,Meng20}, thus limiting their efficiency. Other innovations
that do not symmetrize the meander geometry have focused on high-refractive
index dielectric media surrounding the nanowires to reduce the effective
grating-index contrast
\cite{Redaelli_2016,Zheng_2016,Redaelli_2017,Xu_2017,Mukhtarova_2018}.
Alternatively, instead of minimizing PS, deliberate introduction of
grating-like asymmetries in the optical stack using dielectric or metal
slots to maximize PS have also been considered \cite{Xu_2018,Li_2019}.

\begin{figure*}[t!]
\centering
\includegraphics[width=0.95\linewidth]{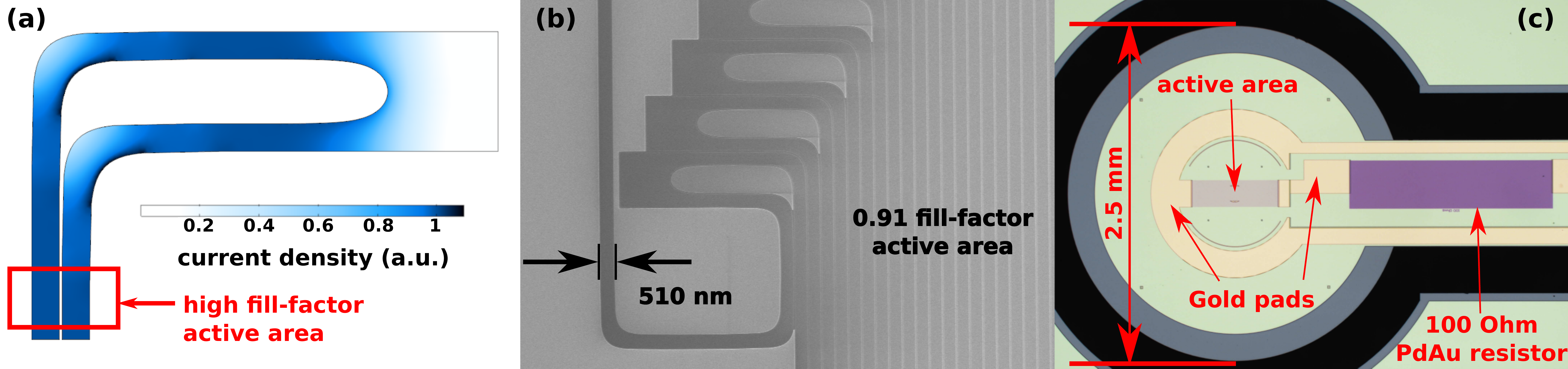}
\caption{\label{fig01} (a) Simulated current density in a candelabra-style hair-pin bend \cite{phidl2021} (see supplementary). (b) SEM image of section of cadelabra meander nanowire showcasing high fill-factor. (c) Optical microgram of device chip showing speed-up PdAu resistor.}
\end{figure*}

The year 2020 witnessed several observations of micrometer-wide
superconducting structures being sensitive to single photons when current
biased. While such scales were trivial for higher-energy photons such as
X-rays \cite{Yang2021}, Korneeva et al. showed the first such instance for
near-infrared (IR) photons \cite{Korneeva2020}. Specifically, they observed
that 3.3 nm thick, 2 \(\mu\text{m}\) wide molybdenum silicide (MoSi)
microstrips could detect photons of wavelength 1 \(\mu\text{m}\). Chiles et
al. modified their tungsten silicide (WSi) recipe to increase the
stoichiometric proportion of silicon, and demonstrated near-IR photon
sensitivity in wires as wide as 4 \(\mu\text{m}\) \cite{Chiles2020}. Similar
results have been observed in niobium nitride (NbN) \cite{Vodolazov2020}.
This new result has spurred interest in gaining a better understanding of
the photon-detection mechanism in such devices. It has also enabled a new
class of superconducting microwire single-photon detectors (SMSPDs),
resulting in new active-area records being set in free-space coupled
devices \cite{Charaev2020,Lita2021,Steinhauer2021}.

Superconducting microwires have already been used to make spiral-meander
SMSPDs by Xu et al. \cite{Xu2021}, acheiving a PS of 1.03 with 92.2\%
efficiency at the wavelength 1550 nm. The maximum fill factor they reported
was 0.8. The presence of curvature in the current's path in the active area
is still suboptimal for SDE due to current crowding
\cite{Yang2009,Clem_2011}. The traditional meander design (parallel strips
of superconductor separated by dielectric medium) when conjoined with
micrometer-scale wire widths offers a trivial means of reaching near-unity
fill factors. This would compensate for the refractive-index grating effect
that differentially scatters orthogonal polarizations
\cite{Anant_2008,Redaelli_2016,Zheng_2016,Redaelli_2017}. The minimum gap
between parallel, straight segments of superconductor in a traditional
meander is limited by the resolution of the electron-beam writing and etch
process, and is typically on the order of 40-100 nm. At such gap widths,
the current crowding at the inner radii of the hairpin bends
\cite{Clem_2011,Baghdadi2021,Xiong2021} of a traditional meander would be
exacerbated for microwires, causing such a device to latch at a very low
bias current \cite{Yang2009}. The current-crowding effect is nullified if
the fill-factor at a hairpin bend is at or below 0.33
\cite{Clem_2011,jonsson2021}.

In this work, we introduce the candelabra meander (see supplementary),
which utilizes optimized 90-degree and 180-degree bending primitives
(defined in the python CAD-layout library \texttt{phidl} \cite{phidl2021})
to slowly turn the microwire outside of the active area, enabling us to
maintain a high active-area fill factor whilst minimizing current crowding
(see fig. \ref{fig01}(a)). The design is inspired by similar structures
used in optical waveguides, where a specific length is to be maintained
within an area/footprint constraint while minimizing optical loss at the
bends. This solution has recently been independently proposed by
J{\"o}nsson et al. \cite{jonsson2021}. The candelabra meander requires a
longer length of microwire to cover the same active area as a traditional
meander (see fig. \ref{fig01}(b)), resulting in increased kinetic
inductance. Using the silicon-rich tungsten silicide (WSi) recipe developed
by Chiles et al. \cite{Chiles2020}, we fabricated fiber-coupled,
candelabra-meander SMSPDs with 0.51 \(\mu\text{m}\) wide wires and a 0.91
fill factor in the active area. These meanders were embedded in the
Bragg-grating and three-layer anti-reflection-coating based optical stack
which was previously employed to break the SDE record \cite{reddy2020}. We
show that these devices have a near-unity PS of better than 1.02 and a high
SDE of better than 91.8\% (67\% confidence at \(2\times10^5\) counts per second)
over a wide bandwidth of 40 nm centered at a wavelength of 1550 nm, and
SDEs in the range of 96.5-96.9\% $\pm$ 0.5\% (at \(1\times10^5\) counts per
second) at 1550 nm. This paves the way for utilization of superconducting
microwires for lowering polarization sensitivity in highly-efficient
single-photon detectors.

\section{Fabrication and experimental setup}
\label{sec:orga02a1b4}

The SMSPDs presented here were fabricated on a 76.2 mm diameter silicon
wafer. Thirteen alternating layers of silicon dioxide (SiO\textsubscript{2}, thickness
\(266.75 \pm 0.84\) nm) and amorphous silicon (\(\alpha\)\text{Si}, thickness \(141.7 \pm
0.27\) nm)--starting with SiO\textsubscript{2}--were deposited onto the substrate using
plasma-enhanced chemical vapor deposition (PECVD), forming a 6.5-period
Bragg reflector at 1550 nm. We then deposited gold terminals and 100 \(\Omega\)
palladium-gold (PdAu) speed-up resistors \cite{reddy2020} (see fig.
\ref{fig01}(c)) using a photolithographic lift-off process. We then
deposited a 2.2 nm layer of silicon-rich WSi \cite{Chiles2020} with a 2-nm
thick \(\alpha\)\text{Si} capping layer using a magnetron sputtering tool. A
candelabra meander for \(0.51\) \(\mu\text{m}\) wide wires and \(50\) nm gap width
was then patterned onto an electron-beam resist layer.

Due to the ultra-thin nature of the WSi layer (which limits amount of light
absorption per transmissive pass), we needed to cover a larger active area
than in comparable optical stacks that utilize other materials and
thicknesses \cite{reddy2020} to account for the extra beam divergence.
Therefore, the candelabra meander covered a rhomboidal active area with a
minor diagonal of length \(80\) \(\mu\text{m}\) and a major diagonal of length
\(174\) \(\mu\text{m}\) (the shortest possible major diagonal for a given minor
diagonal length, fill factor and bend radius). The meander pattern was then
transferred onto the WSi layer using SF\textsubscript{6}-based reactive-ion etching. We
then deposited a three-layer anti-reflection (AR) coating of \(\alpha
\text{Si}\) (\(28\pm0.27\) nm), SiO\textsubscript{2} (\(123.1\pm0.84\) nm), and
\(\alpha\text{Si}\) (\(183.5\pm0.27\) nm) in that order onto the microwire
layer. These thicknesses were determined to minimize reflection of
vertically incident 1550 nm light using rigorous coupled-wave analysis
(RCWA) simulations \cite{Moharam_1981,Li_2016}. Vias were then selectively
etched into the AR-coating to enable wirebonding access to the gold pads.
Deep-reactive-ion etching was then used to etch through the wafer substrate
in a keyhole pattern (see fig. \ref{fig01}(c)), which enabled easy
liberation of the device dies and their mounting into the fiber-ferrule
self-aligning package developed by Miller et al. in 2011 \cite{Miller_2011}.
\texttt{SMF28e+} fiber pigtails terminating at AR-coated, 2.5-mm-diameter ceramic
ferrules were then inserted into the self-aligning packages.

\begin{figure}[h!]
\centering
\includegraphics[width=\linewidth]{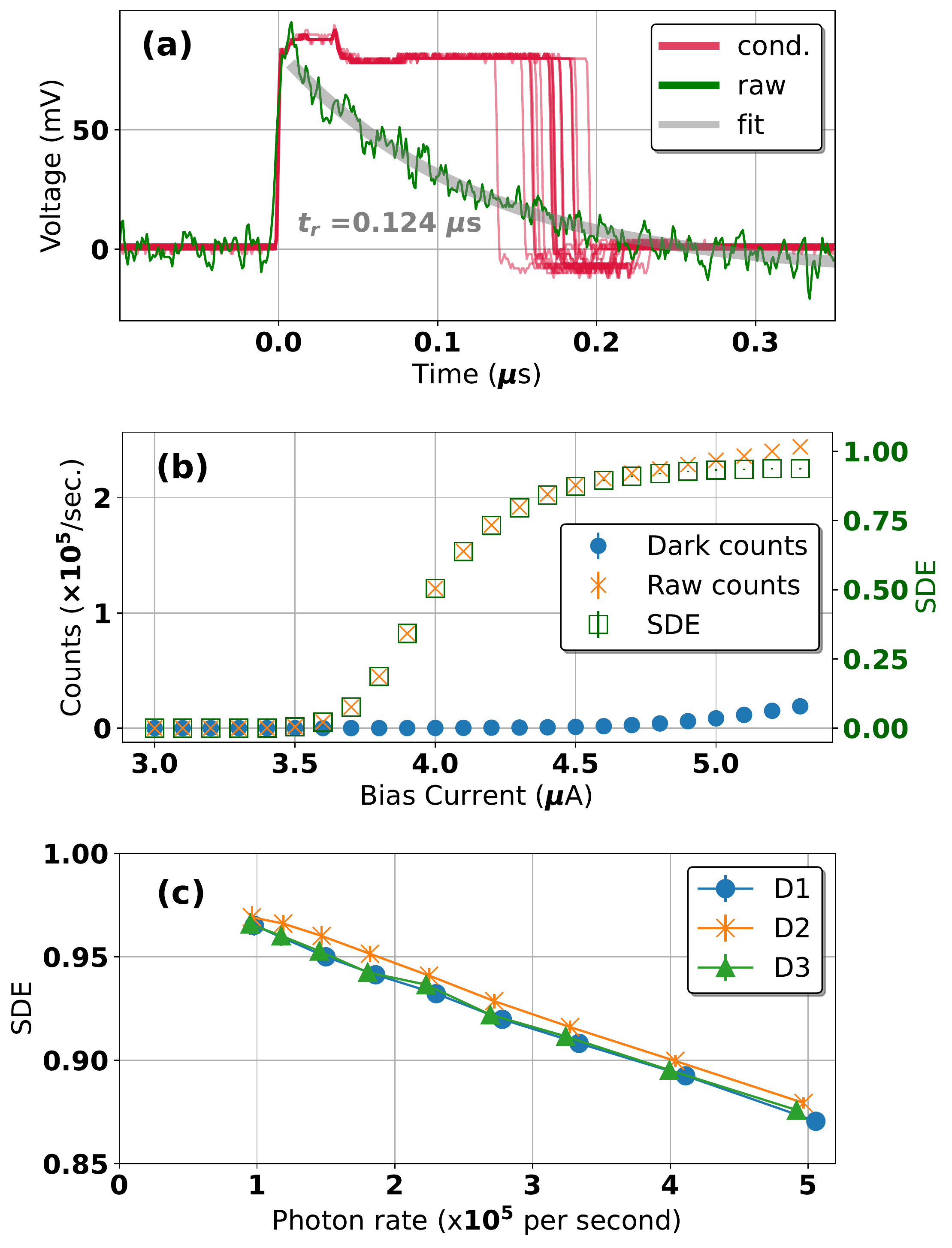}
\caption{\label{fig02_3panel} (a) The raw amplified RF-pulse from device D1, as well as twenty comparator-conditioned pulses vertically scaled-down by a factor of 10. (b) The counts versus bias-current curve at 1550 nm for device D1 at about \(2.3\times 10^5\) detections per second. (c) The system-detection efficiencies (SDE) for various incident photon rates for all three devices biased at 5 \(\mu\)\text{A}, for 1550 nm photons.}
\end{figure}

Four devices from a single wafer were mounted inside a sorption-based
cryostat and cooled to 720-730 mK. The devices were electrically accessible
through SMA ports, and optically accessible through splicing into the bare
ends of the fiber pigtails outside of the cryostat. The system-detection
efficiency (SDE) reported here is defined as the probability for the device
to register a detection given that a photon is launched into the fiber
pigtail \cite{reddy2020}. All measurements were performed using a highly
attenuated, tunable, continuous-wave laser passed through a 1x2 optical
switch and two different types of polarization controllers. An all-fiber
polarization controller was used for algorithmic polarization optimizations
at various wavelengths. A free-space polarization controller was later used
to fully scan the Bloch sphere at 1550 nm. A NIST-calibrated power meter,
and a rack-mounted, ``monitoring'' power meter were used for all equipment
calibrations \cite{reddy2020}. The devices were quasi-current-biased using a
bias tee, a 100 k\(\Omega\) series resistor, and a programmable voltage
source. The detection pulses were amplified using two room-temperature RF
amplifiers, conditioned into square pulses using comparators, and plugged
into an electronic pulse counter. The design, fabrication, calibration
procedures, and error analysis are described in greater detail in the
supplementary material.

\section{Measurement results and discussion}
\label{sec:orgd0e69c1}

\begin{figure}[h!]
\centering
\includegraphics[width=\linewidth]{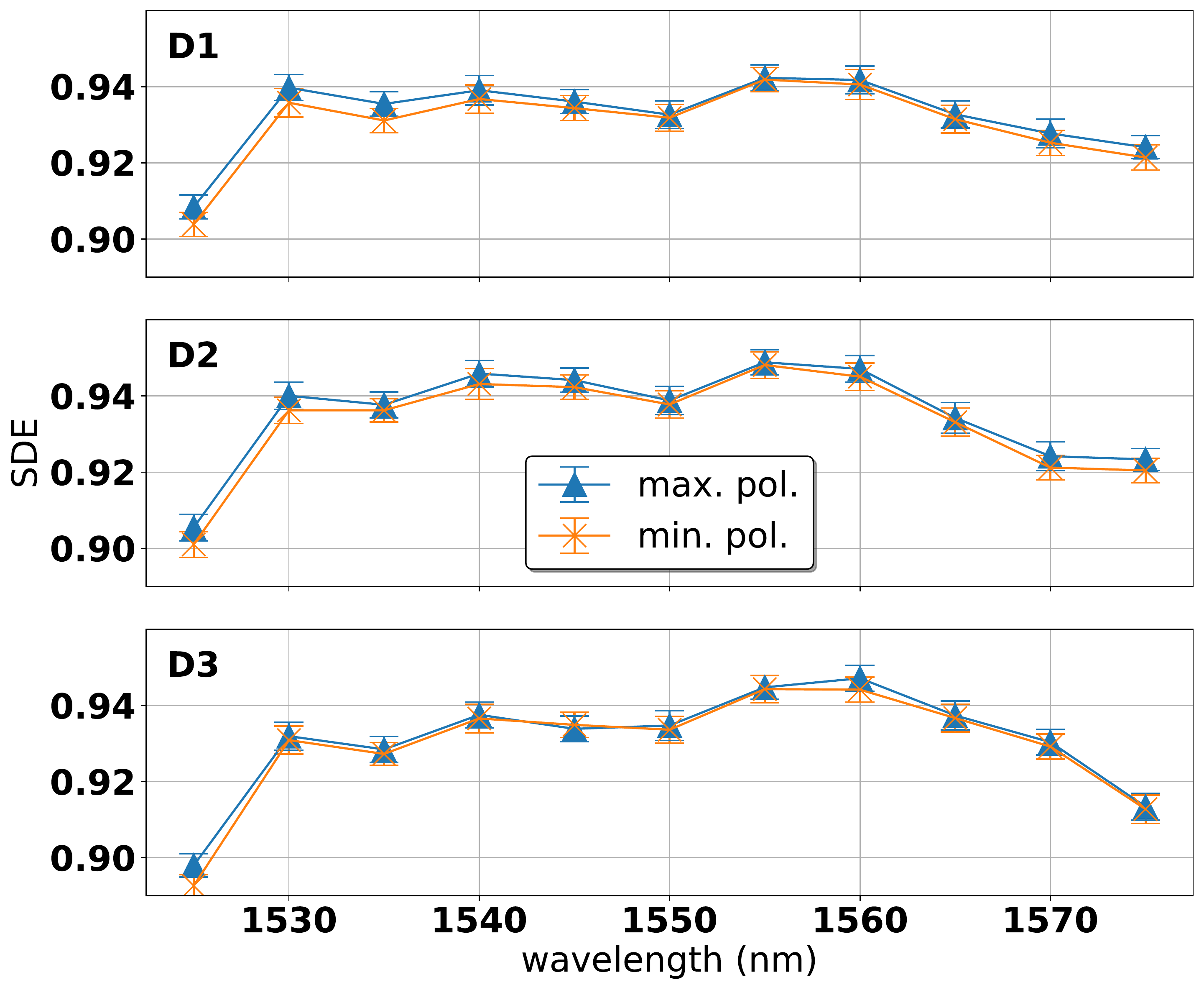}
\caption{\label{fig03_BB} SDE versus wavelength for all three detectors (D1, D2, D3) at count-rates of \(2\times 10^5\) per second. The rates were maximized and minimized (see legend) at constant incident photon rate using an all-fiber polarization controller and the \texttt{python} \texttt{nlopt} library.}
\end{figure}

\begin{figure*}[t!]
\centering
\includegraphics[width=0.95\linewidth]{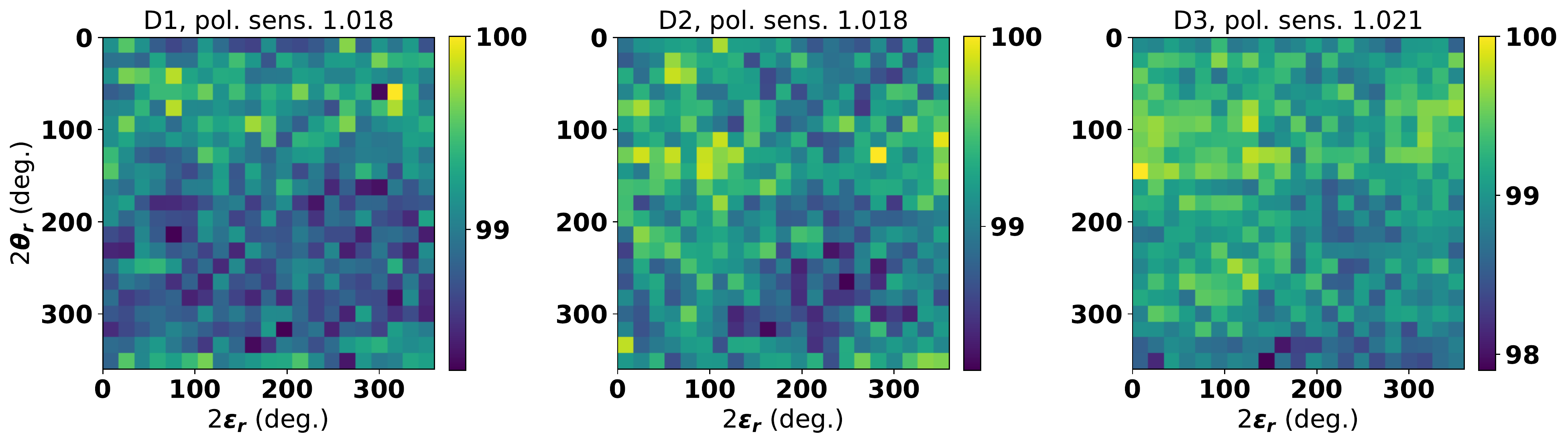}
\caption{\label{fig04_pol} Normalized count rates at a constant photon rate (of about $2.3\times 10^5$ per second) with polarization state varied over the entire Bloch sphere using a free-space polarization controller. See subplot titles for detector numbers.}
\end{figure*}

One out of the four mounted devices was found to be electrically open when
cooled down. We therefore only present the results for the remaining three
devices, labeled D1, D2, and D3. The kinetic inductance of 2.2-nm thick,
Si-rich WSi was measured to be \(275\) \(\text{pH}/\text{sq.}\), which is
nearly thrice the typical value for SNSPDs. This, along with the bigger
active area required, counteracts the gains made in decreasing total
inductance through micrometer-scale wire widths. We fabricated a 100
\(\Omega\) speed-up resistor in series with the microwire to gain a factor of
two in recovery time \cite{reddy2020}, yielding a value of \(\approx 120\) ns.
The width of the comparator-conditioned square pulses averaged to around
\(175\) ns (see fig. \ref{fig02_3panel}(a)). The square pulses showed high
variance in temporal width due to electronic noise affecting the
hysteresis-voltage threshold trigger set at the comparators. Figure
\ref{fig02_3panel}(b) shows the counts versus current bias for detector D1
at \(2.3\times10^5\) counts per second and 1550 nm photon wavelength. All
three detectors showed internal saturation at such count rates, with a
dark-count rate of \(10^4\) per second when biased at 5 \(\mu\text{A}\).

Figure \ref{fig02_3panel}(c) shows the measured SDE at optimal polarization
(optimized using the all-fiber polarization controller and the \texttt{python}
\texttt{nlopt} library) versus 1550 nm photon rate for all three devices. The
pile-up effect resulting from the \(\approx 175\) ns conditioned-square-pulse
reset time \cite{Liu2019,Li:19}, along with any residual device nonlinearity,
results in a detection-rate dependent SDE. The standard error bars on the
SDE estimate are $\pm$(0.39-0.42) \% at a photon rate of \(2.3\times10^5\) per
second, and $\pm$(0.49-0.52) \% at photon rates of \(1\times10^5\) per second
(see supplementary material). The SDE at photon rates of \(10^5\) per second
are around 96.5-96.9 \% across the three devices, which is comparable to
high-efficiency SNSPDs with 100 ns recovery times \cite{reddy2020}. The SDE
vs. photon rate trend line indicates that these devices are asymptotically
fully efficient at ultra-low photon rates, and that no light is being lost
due to beam divergence. We designate a rate of \(2\times10^5\) per second as
a conservative, dominant regime of application, and report efficiencies and
polarization sensitivities at these light levels in the abstract and
conclusion of this document. Furthermore, we report all efficiencies at a
bias current of \(5\) \(\mu\text{A}\), which is about 94-96\% of the switching
current across all three devices.

In fig. \ref{fig03_BB} we plot the SDE for all three devices measured at a
photon rate of \(2\times10^5\) per second across the wavelength range of
1525-1575 nm. The \texttt{nlopt} \texttt{python} library was used to find
the settings for the all-fiber polarization controller that maximized and
minimized the SDE at a given incident light level. All three detectors
showed mean SDEs greater than 92\% in the 1530-1570 nm wavelength range.
For reference, the maximum possible SDE (limited due to pile-up effect
\cite{Coates1968,Walker2002}) for devices with a dead-time of 175 ns at a
continuous input photon rate of \(2\times10^5\) per second is 96.5\%. This
procedure indicated that the all-fiber-controller-derived PS did not exceed
1.006 across the entire bandwidth measured. The PS in some instances was
measured to be very close to unity, resulting in some difficulty in
optimization for the \texttt{nlopt} library. The optimization step for
device D3 at 1545 nm took nearly half-an-hour to halt for both maximization
and minimization, resulting in a ``min. pol.'' mean-SDE value exceeding the
``max. pol.'' mean-SDE value.

The all-fiber polarization controller is not guaranteed to sample the
entire space of polarization states. Therefore we replaced it with a
free-space polarization controller, which transmits the light in free-space
through a linear polarizer, a quarter-wave plate, and a half-wave plate,
all three of which are mounted on controllable rotary mounts in that order.
This controller was used to scan the entire Bloch-sphere of polarization
states. Figure \ref{fig04_pol} shows plots for transmission-corrected (see
supplementary material) counts normalized to the maximum counts across
\(21\times21\) equally spaced polarization settings on the Bloch sphere for
all three detectors. The counts were measured at an average count rate of
\(2.3\times10^5\) per second while the detectors were biased at 5
\(\mu\text{A}\), and the measurement took 20 minutes for each device. Both
dark counts and laser power had to be monitored at each polarzation
setting. A further 20 minutes was required after each measurement session
(per device) to measure the free-space polarization controller transmission
correction using two power meters at classical light levels (see
supplementary material). This measurement yielded PS of \(1.018-1.021\pm
0.008\) for the three devices without any smoothing function applied to the
plotted data. We report a conservative value of \(1.02\pm 0.008\) for PS for
our devices in the abstract and conclusion of this document.

The microwire recipe used in these devices \cite{Chiles2020} required the
superconducting layer to be ultra-thin at around 2.2 nm. This is thinner
than typical for microwire devices explored thus far
\cite{Korneeva2020,Vodolazov2020,Charaev2020,Xu2021}. This resulted in a
larger active-area requirement, causing a large kinetic inductance. We
employed a speed-up resistor to improve the recovery time. This, along with
the substantial length of the candelabra meander, resulted in a large
timing jitter of 1.5 ns. The candelabra meander, when used in conjunction
with superconducting microwires, can trivially hit near-unity PS values
\cite{Meng20} due to their large fill-factors in the active area.
Additionally, the absence of bends within the active area can ensure that
the microwires are photon sensitive across their entire lateral width,
enabling simultaneous near-unity-PS and high-SDE single-photon detection
across a wide range of wavelengths. This capability will prove fruitful for
quantum optics experiments involving wavelength-division multiplexing, or
time-frequency entanglement spanning the low-loss C-band from fiber-optical
communications.

\section{Conclusion}
\label{sec:orgdd61f1b}

We introduced the candelabra meander as a new geometry for superconducting
nanowire and microwire single-photon detectors. This meander enables
high-fill factors in the active area without the deleterious effects of
current crowding at the hairpin bends that plagued the traditional meander
geometry. We utilized this in the fabrication of WSi-based SMSPDs with 0.51
\(\mu\text{m}\) wide microwires and a fill factor of 0.91 in the active area.
We embedded the SMSPDs in the Bragg-reflector-based optical stack optimized
for high efficiency detection of 1550 nm photons. We then fiber-packaged
these devices and measured their polarization sensitivities and
system-detection efficiencies at various wavelengths and photon rates in
the near-IR region. We showed that this design achieves a PS of \(1.02\pm
0.008\) and high efficiencies of greater than 92\% across a 40 nm bandwidth
centered at 1550 nm. This furthers the goal of development of fiber-coupled
single-photon detectors with joint high performance for multiple measures.

\section{Data availability statement}
\label{sec:orgaea33c8}

All of the experimental data gathered during the measurements made for
these results are available on a \texttt{Zenodo} repository:
\url{https://doi.org/10.5281/zenodo.6036210}. Copies of the same can be provided
by the authors upon request. The data and the python code for processing
and plotting will be provided as a \texttt{zip}-archive file (\texttt{SHA-1}
\texttt{91599a67964d30b695b50b4719985533e98869a0}) with a digital signature from
the primary author (\texttt{gpg} fingerprint: \texttt{CC49 3CCF 1104 0DC7 36C4 5A7C E4E5
0022 5ED1 7577}).

\section{Acknowledgments}
\label{sec:orgf2ba722}

The authors would like to acknowledge Igor Veyshenker for providing us with
power-meter calibration. We thank Dr. Gautam A. Kavuri for help with
timing-jitter measurements. We thank Prof. Juliet Gopinath and her group
for accommodating our cryostat and experimental setup in their laboratory
space in the EECE department at University of Colorado, Boulder.

%

\end{document}


\maketitle

\section{Device design and fabrication}

\begin{figure}[h!]
\centering
\includegraphics[width=0.8\linewidth]{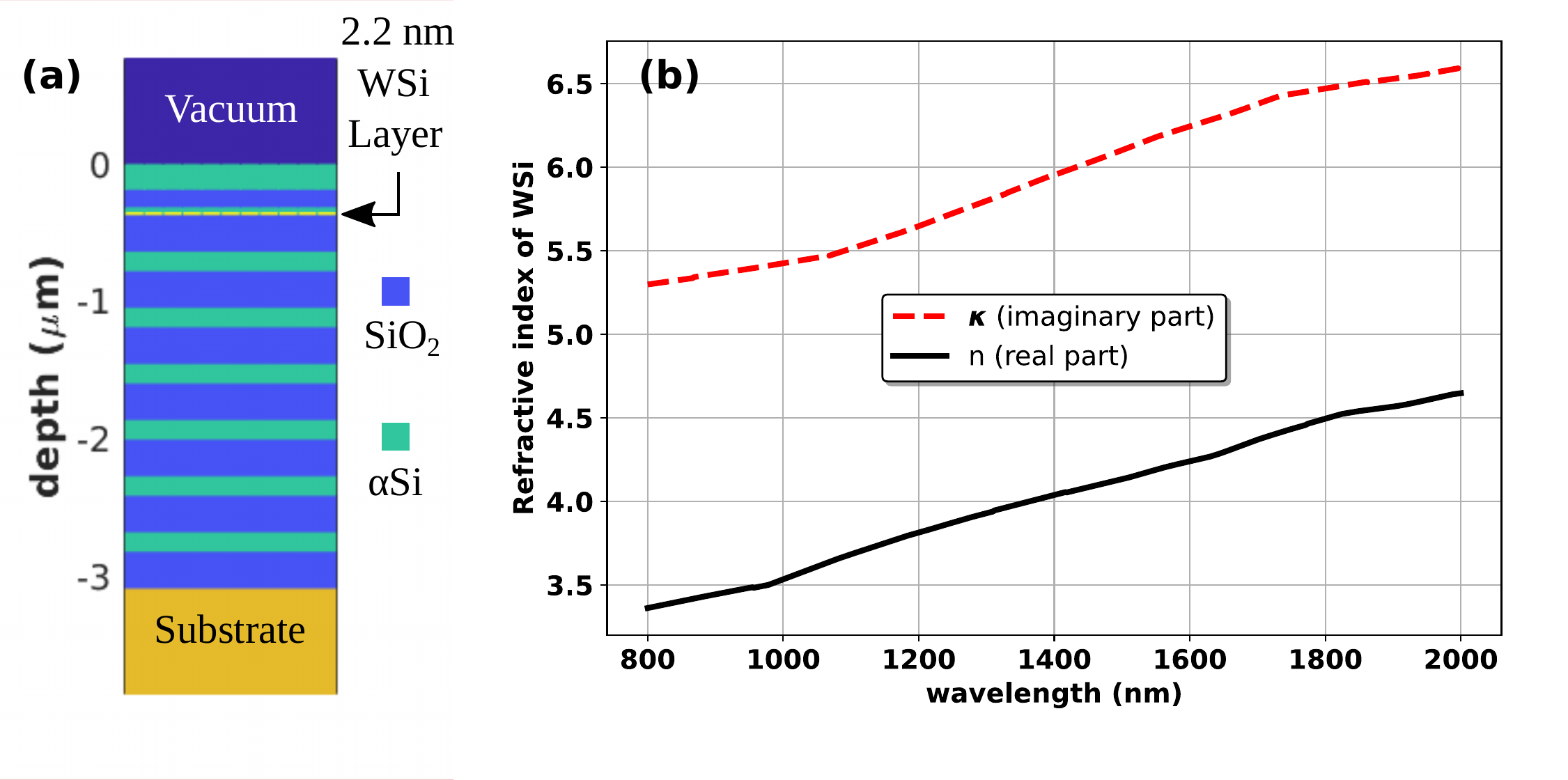}
\caption{\label{fig_supp1} (a) Vertical slice of the Bragg-reflector-based
  optical stack. The microwire is buried beneath a three-layer
  antireflection coating. Light is incident from the top. (b) The
  refractive index estimate of the 2.2 nm thick WSi layer, estimated from
  transmittance and reflectance measurements from a sputtered deposit on a
  UV-fused silica wafer.}
\end{figure}

The Bragg-reflector based optical stack \cite{reddy2020} consists entirely
of two dielectric materials with an index contrast, and the sputtered
superconducting thin film (with an ultra-thin capping layer of amorphous
silicon to prevent oxidation). We deposited all of our dielectric layers in
both the Bragg reflector and the antireflection coating using a
Plasma-enhanced chemical vapor deposition (PECVD) tool. The materials used
are silicon dioxide (SiO$_2$) and amorphous silicon ($\alpha\text{Si}$). Their
refractive indices were measured as $1.453\pm 0.0005$ (SiO$_2$) and
$2.735\pm 0.0005$ ($\alpha\text{Si}$) at a wavelength of $1550$ nm. The
refractive index of the 2.2 nm thick tungsten silicide (WSi) layer was
estimated to be $(n, \kappa) = (4.19, 6.18)$ (see fig. \ref{fig_supp1}(b)) using
transmittance and reflectance measurements off of a deposited layer on
UV-fused silica.

With an ideal Bragg reflector composed of thirteen alternating layers of
SiO$_2$ and Si (starting with SiO$_2$, see fig. \ref{fig_supp1}(a)) with
optical thicknesses of $\lambda/4$ at 1550 nm, we used rigorous coupled-wave
analysis (RCWA) to optimize a three-layer antireflection (AR) coating to
minimize back-reflection of vertically incident light at 1550 nm. For a
microwire grating of width 0.51 $\mu\text{m}$ and an inter-wire gap of 50 nm,
the optimum AR coating was 28 nm ($\alpha\text{Si}$), 123 nm (SiO$_2$), and
183.5 nm ($\alpha\text{Si}$), from bottom to top.

\begin{figure}[h!]
\centering
\includegraphics[width=0.8\linewidth]{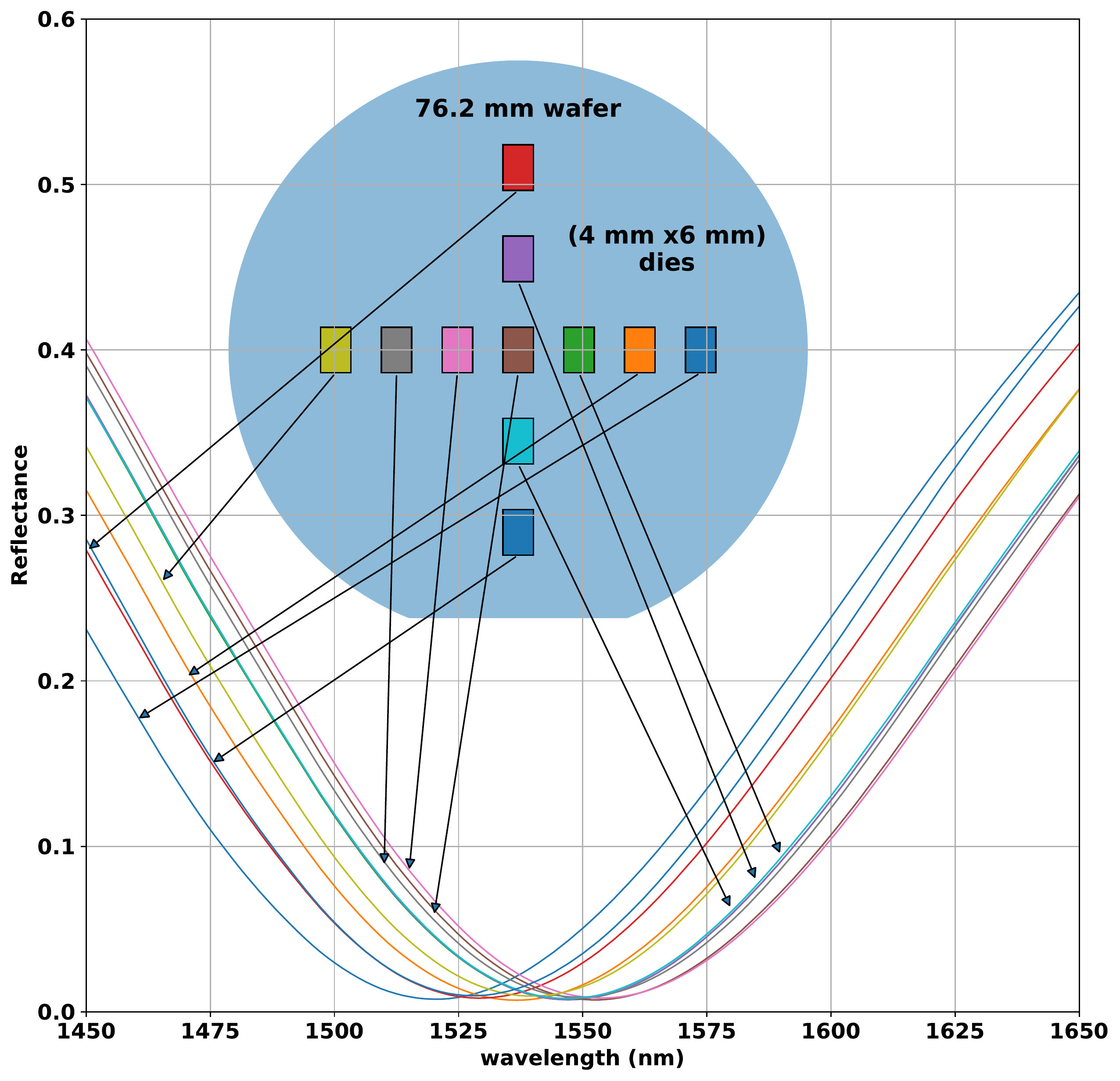}
\caption{\label{fig_supp2} Reflectance of a vertically incident 10
  $\mu\text{m}$ spot focused on the top surface of the devices (measured
  using a Filmetrics reflectometry tool). The plots for various die
  locations on the 76.2 mm wafer indicate that the devices on the wafer's
  periphery ended up optimized for absorption at shorter wavelengths.}
\end{figure}

The layer-thickness errors at the center of a 76.2 mm diameter substrate
wafers were sub-1 nm between deposition runs on the tool. However, the
layer thicknesses were not uniform across the wafer due to the tool's
inherent design. The three-layer AR coating optimum was sought in a
parameter regime where increase in the top-most Si-layer thickness moved
the minimum-reflectance wavelength higher with negligible change in the
reflectance at said wavelength. Therefore, during fabrication, we
deliberately undershot the thickness of the final AR-coating layer. This
allowed us to repeatedly transfer the wafer in the cleanroom between the
PECVD tool and a Filmetrics reflectometry tool. The reflectance of a 10
$\mu\text{m}$ spot focussed on the top surface of the vertical stack was
sequentially monitored as short, corrective depositions to the top layer
were made. This method allowed for easy compensation for all of the
collective fabrication errors that may have occurred in all the previous
steps. However, due to the nonuniformity of the dielectric layer
thicknesses across the wafer, the optimum wavelength for different devices
depended on the die location on the wafer (see fig. \ref{fig_supp2}). Our
devices were placed on 4 mm $\times$ 6 mm dies, and were optimized for 1550 nm
only near the center of the wafer.

The candelabra-meander pattern was drawn using the \texttt{python}
CAD-layout library \texttt{phidl} \cite{phidl2021}. A superconducting wire
``goes normal'' when the current density exceeds a certain value related to
its depairing current. When a conductive path in any current-biased thin
film bends, the current crowds towards the inner radius (see main
manuscript), meaning that the current density is no longer uniform across
the width of the conductive path like in its straight segments. The current
density at the inner radius can exceed the current density in the straight
segments by an amount related to the width of the path and the bend radius.
In superconducting wires, these regions can also be a significant source of
dark counts \cite{Baghdadi2021}.

\begin{figure}[h!]
\centering
\includegraphics[width=\linewidth]{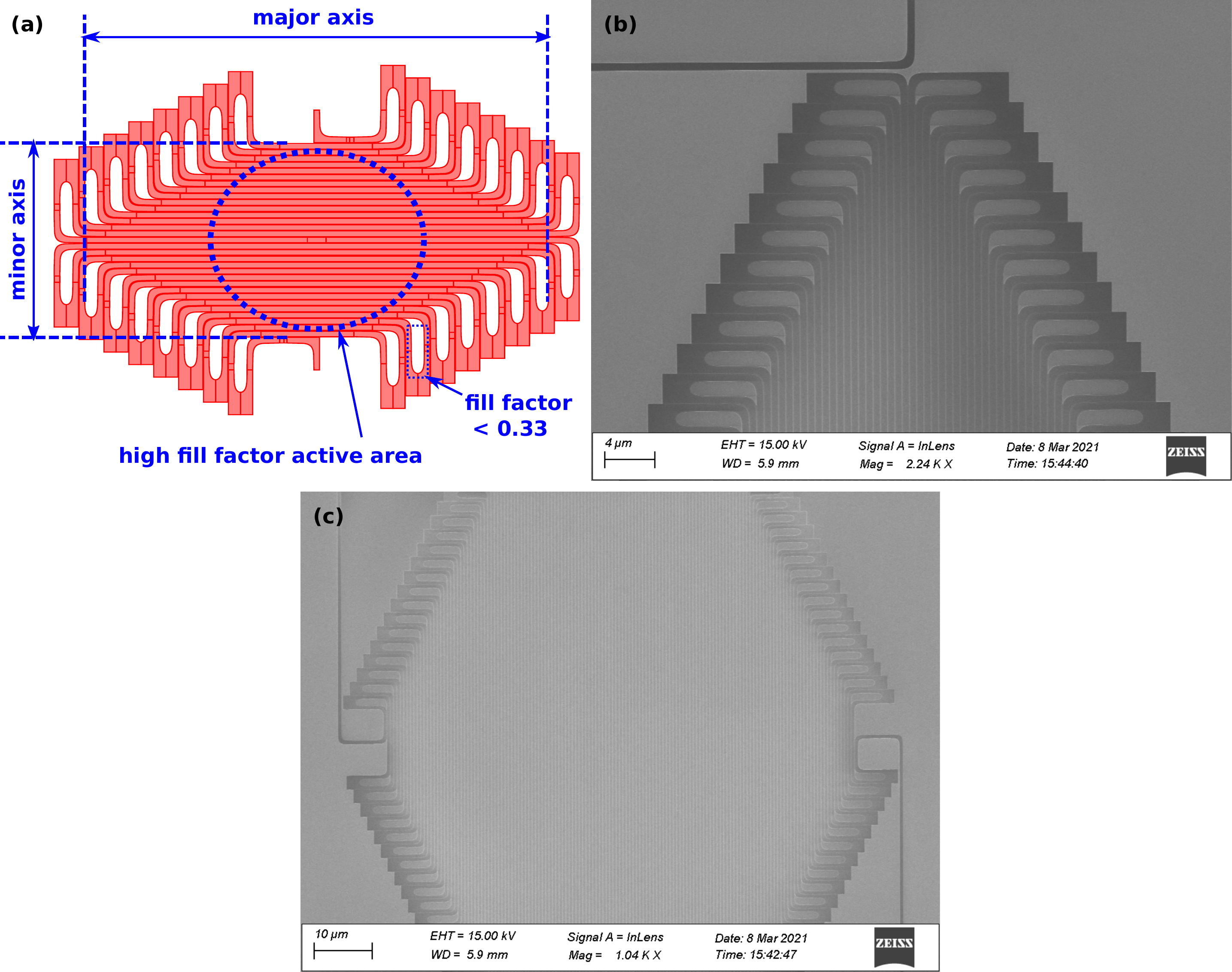}
\caption{\label{fig_supp3} (a) A schematic of the candelabra meander
  \cite{phidl2021}. (b, c) SEM images of the candelabra meander etched into
  WSi.}
\end{figure}

For a hairpin bend, where in the direction of the path turns by 180
degrees, for the current density at the bend to not exceed that in the
straight segments, the bend-geometry must match the current-distribution
contour lines, and the local fill factor must be $\le 0.33$
\cite{Clem_2011,jonsson2021}. The \texttt{phidl} library contains built-in
methods for plotting optimized 90-degree and 180-degree turns for any given
wire width. These were utilized to create the candelabra meander (see fig.
\ref{fig_supp3}(a)), which allows for high fill factors in a designated
active area by moving all the turns and low-fill-factor hairpin bends to
the outside of it \cite{jonsson2021}. This pattern was written into a PMMA
resist coating using an e-beam writer. It was developed in cold
($5^\circ\text{C}$) 1:3 MIBK:IPA solution, and the pattern was then etched into
the WSi layer using an SF$_6$ -based reactive-ion etch recipe. Figures
\ref{fig_supp3}(b) and \ref{fig_supp3}(c) show SEM images of the pattern
after it has been etched into WSi.

All the gold pads (50 nm electrical terminals, and 100 nm spacer layer) and
the palladium-gold speed-up resistor \cite{reddy2020} were deposited using
an LOR3A+SPR660 photolithographic lift-off process. They were all
sandwiched between 2 nm layers of titanum for adhesion. Two 100-nm thick
gold spacers on top of the stack on either side of the active area
(separated by 120 $\mu\text{m}$) prevented the fiber-pigtail/ceramic-ferrule
front faces from making contact and damaging the optical stack in the
self-aligning fiber package \cite{Miller_2011}. This ensured the presence
of vacuum/air between the optical stack and the AR-coating on the fiber
face.

\section{Efficiency measurement procedure and error analysis}

With multiple groups now reporting near-unity detection efficiencies for
SNSPDs/SMSPDs \cite{reddy2020,Hu2020,Chang2021}, the need has arisen to
standardize and tighten the procedures for the same. While there are
multiple approaches to estimating detector efficiency
\cite{Marsili_2013,Gerrits2019}, they all rely on one or more power meters
which are well calibrated to report accurate readings at known wavelengths,
power-meter range settings, and light levels \cite{Vayshenker_2006}. The
method we have developed and present in detail here requires two power
meters with low electrical noise. Only one of these are required to be
calibrated for accurate readings at a single classical light level and
range (gain) setting at all desired wavelengths. This method was
developed specifically for accessibility and ease of implementation.

Our procedure estimates the SDE as the ratio of the time-averaged number of
detection events registered from the device to the expected count rate,
with light from a highly-attenuated continuous-wave laser as the input. The
expected count rate is calculated by measuring the laser power without
optical attenuation using a calibrated power meter, and then estimating the
attenuation applied. Optical attenuation estimation requires power
measurements both with and without the attenuation applied. No power meter
designed for classical light-levels is sensitive to light at the low powers
necessary for testing single-photon detectors. Therefore, multiple modular
optical attenuators are used, with each one applying a fraction of the
required total attenuation. Calibrating these modular attenuators
individually involves measuring and comparing optical powers that are
several orders of magnitude apart. The power meter used for this purpose
will need to operate at different range/gain settings to measure and
compare such optical powers. Both the nonlinear response of this power
meter within a range setting, as well as response discontinuities across
contiguous range-setting changes, are estimated using a
nonlinearity-correction step. And lastly, an optical switch in the system
is also calibrated using two power meters in the same range setting, one of
which has been calibrated in that range setting to act as an absolute power
reference.

The chief advantage of this procedure is that it only requires one optical
element (the calibrated power meter) to have been externally characterized
at one classical light level (and range setting) to act as a reference. All
the other optical elements (switch, attenuators, other power meters) are
calibrated live. Additionally, the same experimental setup serves for both
device SDE measurement and optical-element calibration. The only physical
modification to the setup required during the procedure is the
fiber-splicing involved in routing the prepared light to the device under
test (DUT) instead of the calibrated power meter.

\begin{figure}[h!]
\centering
\includegraphics[width=0.8\linewidth]{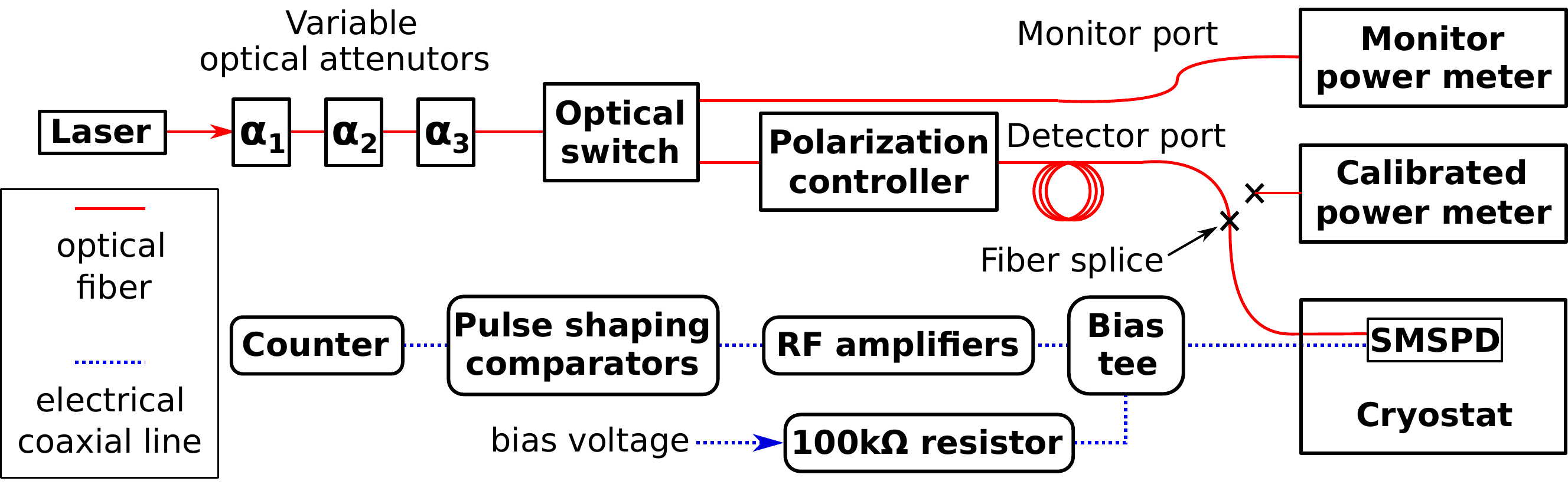}
\caption{\label{fig_supp4} SDE measurement setup, illustrating the optical
  calibration and input state preparation, as well as the electrical
  biasing and readout circuitry. SMSPD stands for superconducting microwire
  single-photon detector.}
\end{figure}

The full schematic of the experimental setup is shown in fig.
\ref{fig_supp4}. The optical components used are a tunable, continuous-wave
laser (\texttt{Ando AQ8201-13B}), three all-fiber tunable optical
attenuators with at least one equiped with a monitoring port (\texttt{Ando
  AQ8201-33(M)}), an optical switch (\texttt{Ando AQ8201--422}), a
monitoring power meter (MPM) (\texttt{Ando AQ8201-22}), a NIST-calibrated
power meter (CPM) (\texttt{Agilent HP81533B} with \texttt{InGaAs} optical
head), and a programmable polarization controller. For efficiency
measurements, a \texttt{FiberControl MCP-01} all-fiber polarization
controller was used. For sampling the Bloch sphere for all input
polarization states, it was swapped with a free-space
\texttt{Hewlett-Packard HP8169A} polarization controller. The laser,
attenuators, optical switch, and the MPM were all slotted into an
\texttt{Andorack} \texttt{AQ8201A} mainframe with a \texttt{GPIB} control
interface. They are all daisy-chained optically using black-jacketed
\texttt{FC/APC} patch chords. The CPM is ``perpetually'' attached to an
\texttt{SMF28e+} fiber pigtail with the same AR-coating on its ceramic
front face as the ones inserted into the SMSPDs. This fiber pigtail is
identical to the one used during CPM calibration at NIST.

The two output routes of the optical switch are labeled the monitor port,
and the detector port (see fig. \ref{fig_supp4}). The polarization
controller is attached to the output route meant for the devices under
test. The fiber port at the output of the polarization controller is
labeled the detector port. This fiber port is an exposed, bare
\texttt{SMF28e+} fiber which can be spliced either to a fiber pigtail
plugged into the CPM (as is done during the switch calibration step,
detailed below), or to the fiber optically coupled to the DUT. The system
detection efficiency (SDE) is defined as the probability of the detector
producing a detection event given that a photon is present in the optical
mode exiting the detector port. A bad (lossy) fiber splice into the DUT
fiber will underestimate the SDE. A bad fiber splice into the CPM during
the switch calibration step will overestimate the SDE. Special care is
taken to achieve the best fiber splice into the CPM (through repeated
attempts) for the switch calibration step.

The electrical components in the setup are a low-noise programmable voltage
source (\texttt{Keithley 213}), a pulse counter (\texttt{SRS SR400}), a
digital multimeter (\texttt{Keithley 199}) (not shown) used to monitor the
voltage drop across the DUT, room temperature low-noise
\texttt{Mini-Circuits} RF amplifiers (\texttt{ZFL500LN+},
\texttt{ZFL1000LN+}), a standard \texttt{Mini-Circuits} bias tee, a custom
comparator circuit, and assorted inline resistors and low-pass filters (not
shown). As mentioned in the main manuscript, the DUTs are
quasi-current-biased using the voltage source and a 100 k$\Omega$ series
resistor.

\subsection{Calibrated power meter readings}

\begin{figure}[h!]
\centering
\includegraphics[width=\linewidth]{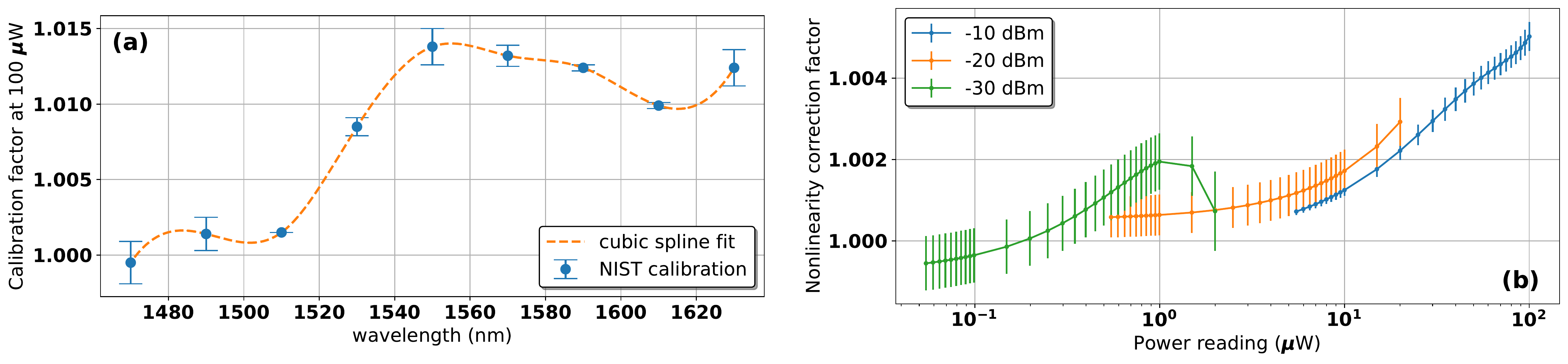}
\caption{\label{fig_supp5} (a) NIST-provided calibration factor
  ($CF_{\text{CPM}}(\lambda)$) for the calibrated power meter at power reading of
  100 $\mu$W. (b) The nonlinearity correction factors ($CF_{\text{NL}}(\lambda, r,
  v)$) versus power-meter reading ($v$) for the monitoring power meter estimated
  for various range settings at $\lambda=1550$ nm.}
\end{figure}

The NIST-calibrated \texttt{InGaAs} optical head and \texttt{HP81533B} pair
were calibrated at specific near-infrared wavelengths at a light level of
100 $\mu$W (see fig. \ref{fig_supp5}(a)). The CPM reading of 100 $\mu$W taken
with range setting of $-10$ dBm is to be divided by the calibration factor
($CF_{\text{CPM}}(\lambda)$) to recover the actual power. Although the company
datasheet lists a standard deviation of 2.5\%, the calibration factors for
this specific CPM were estimated to better than 0.14\% (see errorbars in
fig. \ref{fig_supp5}(a)). Throughout error analysis, we will use the
conservative value of $\sigma_{\text{CPM}}/CF_{\text{CPM}}(\lambda) = 0.14\%$ $\forall \lambda$
for relative standard uncertainty. NIST offers additional types of
calibrations, such as nonlinearity correction for different range settings
and light levels. Our method does not rely on these.

\subsection{Monitoring power meter nonlinearity correction}

The nonlinearity correction factor ($CF_{\text{NL}}(\lambda, r, v)$, where $v$ is
the power-meter reading, $r$ is the range setting, and $\lambda$ the wavelength)
is estimated for the MPM by measuring its power readings in the power range
of $P_{\text{MPM}} = 1 \text{nW}-100 \mu\text{W}$ and all range settings ($r
\in (-10, -20, -30, -40, -50, -60)$ dBm) \cite{Vayshenker_2000}. For this,
two of the attenuators are used (the third one is held at ``0 dB'' setting)
and the optical switch is set to route the light to the MPM. No assumptions
are made about the precise value of the attenuation applied, but attenuator
repeatability is assumed. The procedure for these measurements is shown in
algorithm \ref{alg:nonlin}.

\begin{algorithm}
  \caption{Nonlinearity factor raw power meaurements}\label{alg:nonlin}
  \begin{algorithmic}[1]
    \Procedure{console\_nonlinearity}{$pm,att1,att2,out$}\Comment{$pm\equiv$
      MPM, $att(1,2)\equiv$ attenuators, $out\equiv$ file}
    \State{$N\gets10$}\Comment{Number of reads at each setting}
    \State{$xlist\gets[20, 15]$}
    \State{$xlist.append(arange(10,0.9,-0.5))$}\Comment{10 to 0.9 in
      steps of -0.5}
    \State{$xlist.append(arange(0.95,0.5,-0.5))$}\Comment{0.95 to 0.5 in
      steps of -0.5}
    \State{$base\_array\gets round(10-10*log_{10}(xlist))$}
    \State{$base\_array\gets base\_array-min(base\_array)$}
    \State{$att\_setting\gets\{\}$}\Comment{initialize empty dictionary}
    \For{$rng\in[-10, -20, -30, -40, -50, -60]$}\Comment{$rng\equiv$ range setting}
    \State{$att\_setting[rng]\gets base\_array - (rng + 10) - 3$}\Comment{3 is
      an offset}
    \EndFor
    \State{}
    \For{$rng\in[-10, -20, -30, -40, -50, -60]$}
    \State{$pm.set\_range(rng)$}\Comment{Set MPM range setting to $rng$}
    \State{$pm.zero()$}\Comment{Zero the MPM}
    \For{$\alpha\in att\_setting[rng]$}
    \State{$att1.set\_att(\alpha)$}\Comment{Set attenuation setting on $att1$ to
      $\alpha$}
    \For{$att\_step\in[0, 3]$}
    \State{$att2.set\_att(att\_step)$}\Comment{Set attenation setting on
      $att2$ to $att\_step$}
    \For{$i\gets 1$ to $N$}\Comment{Iterate $N$ times}
    \State{$power\gets pm.get\_power()$}
    \State{$out.write(\alpha, att\_step, rng, power)$}\Comment{Record settings
      and power readings from MPM into file}
    \EndFor
    \EndFor
    \EndFor
    \EndFor
    \EndProcedure
  \end{algorithmic}
\end{algorithm}

\begin{figure}[h!]
\centering
\includegraphics[width=\linewidth]{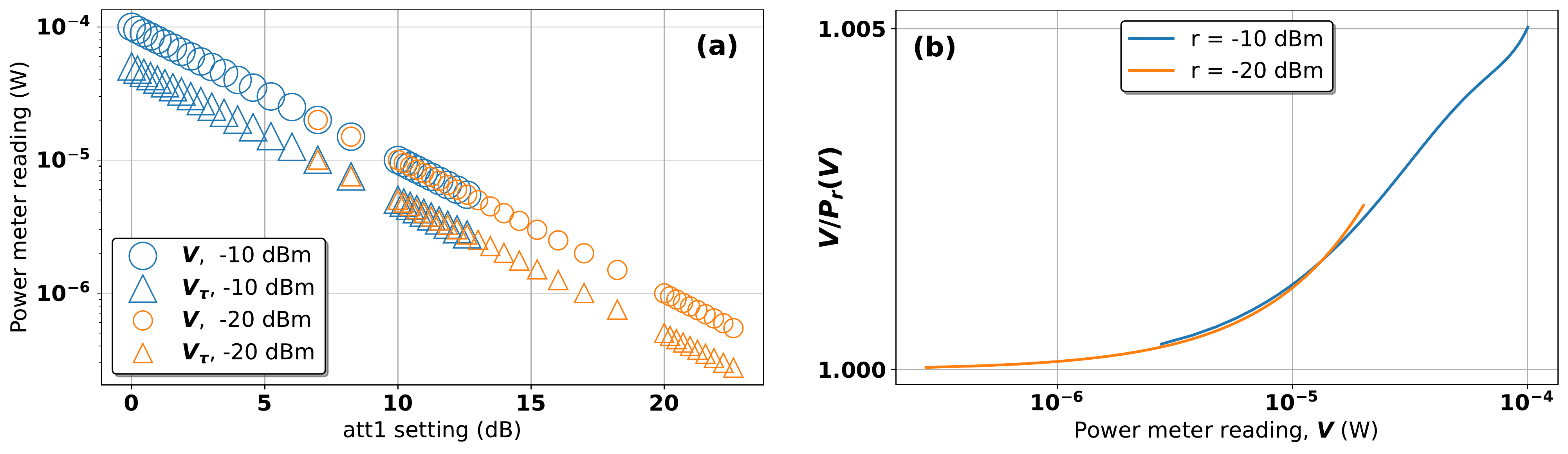}
\caption{\label{fig_supp6} (a) MPM Power readings ($V$, $V_\tau$) under two
  range settings measured at a range of attenuation settings for 1550 nm
  light for nonlinearity calibration (see algorithm \ref{alg:nonlin}). (b)
  The ratio of the readings ($V$) to the polynomial fit ($P_r(V)$) for two
  range settings $r \in \{-10, -20\}$ dBm.}
\end{figure}

Algorithm \ref{alg:nonlin} is run at every desired wavelength. It records
the MPM power readings at various light levels (dictated by the attenuator
settings). Of the two attenuators varied, one ($att1$) is used in a range
of settings to sweep the power reading monotonically, while the other
($att2$) is used as a binary on/off device by switching its settings
between nominal values of ``0 dB'' and ``3 dB''. The actual amount of
optical attenuation applied by the ``3 dB'' setting relative to the ``0
dB'' setting is an unknown (but to be fit) parameter $\tau$. The objective of
algorithm \ref{alg:nonlin} is two-fold, namely: (a) to record an array of
MPM readings over a multitude of $att1$ settings at every MPM range setting
both with $(V_\tau)$ and without $(V)$ the attenuation $\tau$ applied by $att2$,
and (b) to record several MPM readings with the same $att1$ and $att2$
settings across contiguous MPM range settings. Figure \ref{fig_supp6}(a)
shows the readings taken at 1550 nm for two range settings. The points that
overlap in attenuator settings will be used to compute the range
discontinuity.

The nonlinearity correction for the MPM has two components. The first
corrects for the nonlinear response to incident power within a range
setting, and the other corrects for a discontinuity in readings for the
same incident power across a change in range setting. To achieve the
former, we use the \verb|python| \verb|lmfit| library and model the
function that maps the MPM readings (at range setting $r$) to the
linearized power as a polynomial $P_r(V)$, defined as

\begin{equation}
 P_r(V) = V + \sum_{k=2}^{N_r}b_k^{(r)}V^k,
\end{equation}

\noindent where $N_r$ is the order of the polynomial. We can then determine
the coefficients of the polynomial $\{b_k^{(r)}\}$ and the unknown
attenuation $\tau$ by solving for

\begin{equation}
  P_r(V_\tau) - \tau P_r(V) = 0, \forall r \in \{-10, -20, -30, -40, -50, -60\} \text{ dBm},
\end{equation}

\noindent where $V_\tau$ and $V$ are the readings recorded at range-setting
$r$. This polynomial fit is performed by minimizing the residual
simultaneously at all the range settings recorded. The choice of what order
of polynomial to fit for at each range setting ($N_r$) is made by
minimizing the reduced chi-square goodness-of-fit metric over the full
range of choices between 1 and 5 for each range-setting $r$. Figure
\ref{fig_supp6}(b) shows the polynomial fits made using the data in fig.
\ref{fig_supp6}(a). Note the discontinuity across the change in range
setting in the overlapping region.

For the second aspect of nonlinearity correction, namely, range
discontinuity, we consider the MPM readings recorded at the same combined
($att1$, $att2$) attenuation settings across two contiguous range settings
$r$ and $r + 10$ dBm. These readings ($\overline{V}_r$) will be a subset of
the full record of MPM readings from algorithm \ref{alg:nonlin}. These
readings are then used to compute an average ratio of
polynomial-fit-corrected readings across ranges $r$ and $r+10$ dBm, giving
as a value denoted by $RF_{\text NL}(\lambda, r)$. To get the full nonlinearity
correction $CF_{\text NL}(\lambda, r, v)$ for MPM reading $v$ at range setting
$r$, we first commulatively multiply $RF_{\text NL}(\lambda, r')$ for all $r' \in
\{r, r+10, r+20, ..., -20\} \text{ dBm}$, and we then multiply that to
$v/P_r(v)$. More explicitly,

\begin{equation}
  RF_{\text NL}(\lambda, r) =
  \left\langle\frac{P_r(\overline{V}_r)}{P_{r+10}(\overline{V}_{r+10})}\right\rangle_{(att1,
  att2)},\quad CF_{\text
    NL}(\lambda, r, v) = \frac{v}{P_r(v)}\times\left(\prod\limits_{r'=r}^{-20 \text{ dBm}}RF_{\text NL}(\lambda, r')\right).
\end{equation}

The full nonlinearity correction is applied to any MPM reading $v$ at range
setting $r$ by dividing the reading by $CF_{\text NL}(\lambda, r, v)$. The range
setting of $r=-10$ dBm is the highest reached in this experiment, and is
the one at which the calibration factor for the CPM is known. Hence we only
account for range discontinuities up to $r = -10$ dBm ($RF_{\text NL}(\lambda,
r=-10 \text{ dBm}) = 1$). Figure \ref{fig_supp5}(b) plots the nonlinearity
correction factor for the MPM over three range settings for 1550 nm light.
For convenience's sake, we will omit the third argument ($v$) in $CF_{\text
  NL}$ when denoting the nonlinearity correction factor in error analysis.

The standard ($k=1$) uncertainties in the nonlinearity correction factors
for range-setting $-30$ dBm were between $0.061\%$ and $0.075\%$ over the
wavelengths considered. For this error budget calculation, we will use the
conservative maximum value of $\sigma_{\text{NL}}/CF_{\text{NL}}(\lambda, r, v)
=0.075\%$ $\forall \lambda$, $\forall v$ for the relative standard uncertainty.

\subsection{Optical switch calibration}

The optical switch is calibrated at classical light levels of around $100$
$\mu\text{W}$ at desired wavelengths using both the CPM and MPM. The detector
port is spliced into the fiber pigtail coupled to the CPM (see fig.
\ref{fig_supp4}), and the switch is alternatingly ``flipped'' to route the
power to either the monitor port or the detector port while several power
readings are recorded from both power meters. The range settings for both
power meters are set to $-10$ dBm, and the attenuators are all set to a
zero-attenuation setting. This step was carried out with the all-fiber
\texttt{MCP-01} polarization controller only.

\begin{figure}[h!]
\centering
\includegraphics[width=0.5\linewidth]{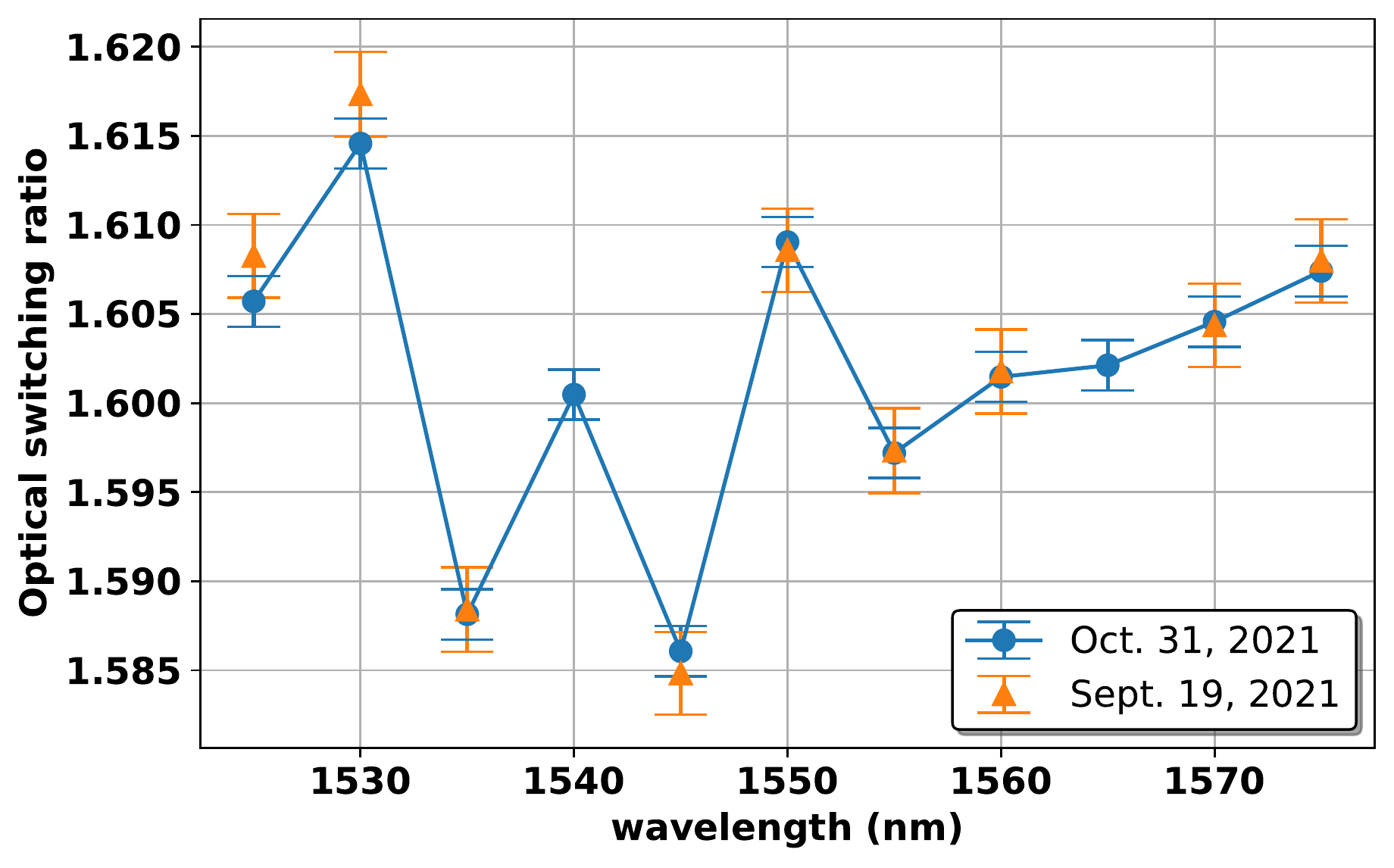}
\caption{\label{fig_supp7} The switching ratio versus wavelength of the
  optical switch measured with classical light levels ($100$ $\mu\text{W}$)
  using the calibrated and the monitor power meters. The switching ratio
  did not change much over 40 days (see legend). The switching ratios used
  for SDE estimates were measured just before and after the efficiency
  measurement runs.}
\end{figure}

The switching ration $R_{\text{SW}}(\lambda) =
P_{\text{CPM}}(\lambda)/P_{\text{MPM}}(\lambda)$ is plotted in fig. \ref{fig_supp7}.
This ratio was calculated before every run which required splicing a fiber
to a DUT. As can be seen in fig. \ref{fig_supp7}, the ratio barely drifts
over 30 days due to the temperature stability of the lab, and the
mechanical stability of the setup. The relative standard uncertainty in the
switching ratio was $\sigma_{\text{SW}}/R_{\text{SW}}(\lambda) \le 0.14\%$ as the MPM
readings were very stable at these light levels.

\subsection{Attenuator calibration}

The optical attenuators were calibrated one at a time, meaning with the
other two attenuators set to zero-attenuation setting. This allows for the
light level reaching the MPM to be large enough for a valid read out with
an appropriate range setting. For any given attenuator setting, the
absolute amount of optical power exiting the detector-port fiber is
$P_{\text{DP}} =
P_{\text{MPM}}/R_{\text{SW}}/CF_{\text{CPM}}(\lambda)/CF_{\text{NL}}(\lambda, r, P_{\text{MPM}})$,
where $P_{\text{MPM}}$ is the power reading from the MPM at wavelength $\lambda$
and range setting $r$. The absolute power is calculated with the
attenuation setting (for the attenuator being calibrated) set to both the
zero-attenuation setting and the setting used in the DUT measurement. The
ratio of the estimates of absolute power is equal to the true attenuation
applied. It is assumed that the attenuations apply cummulatively when all
three attenuators are set to a high-attenuation setting together.

Due to several factor cancellations, the $R_{\text{SW}}$ and
$CF_{\text{CPM}}$ need not be used in attenuator calibration. The relative
standard uncertainty for attenuator $\alpha_i$ is

\begin{equation}
  \frac{\sigma_{\alpha_i}}{\alpha_i}=\sqrt{\left(\frac{\sigma_{\text{MPM}}}{P_{\text{MPM}}}\right)_{r=-10\text{ dBm}}^2+\left(\frac{\sigma_{\text{MPM}}}{P_{\text{MPM}}}\right)_{r=-30\text{ dBm}}^2+\left(\frac{\sigma_{\text{NL}}}{CF_{\text{NL}}(\lambda,
      r=-10\text{ dBm})}\right)^2+\left(\frac{\sigma_{\text{NL}}}{CF_{\text{NL}}(\lambda,
      r=-30\text{ dBm})}\right)^2}.
\end{equation}

For a photon flux of $2\times10^5$ per second, the attenuation setting for all
three attenuators was 31 dB, and the range setting on the MPM during
attenuator calibration was $-30$ dBm, yielding a relative standard
uncertainty of $\sigma_{\alpha_i}/\alpha_i = 0.2\%$.

This procedure (see algorithm \ref{alg:att}) is performed every time the
counts from the DUT are measured for efficiency estimation. The CPM is not
used, and the detector port is spliced into the fiber coupled to the DUT
during attenuator calibration. The optical switch is set to route the light
to the MPM throughout this procedure.

\begin{algorithm}
  \caption{Attenuator calibration}\label{alg:att}
  \begin{algorithmic}[1]
    \Procedure{console\_cal}{$pm,att\_list,attval,rngval,out$}
    \Comment{$pm\equiv$ MPM, $att\_list\equiv$ attenuators,
      $attval\equiv$ attenuator value, $rngval\equiv$ range value, $out\equiv$ file}
    \State{$N=5$}\Comment{Number of measurements}
    \State{$init\_rng = -10$}\Comment{Initial MPM range setting}
    \For{$att\in att\_list$}
    \State{$att.set\_att(0)$}\Comment{Set all attenuators to $0$ dB}
    \EndFor
    \State{}
    \For{$att\in att\_list$}
    \State{$pm.set\_range(init\_rng)$}
    \State{$att\_list.disable()$}\Comment{Disable all attenuators}
    \State{$pm.zero()$}\Comment{Zero the MPM}
    \State{$att\_list.enable()$}\Comment{Enable all attenuators}
    \State{$powers\gets \{\}$}
    \For{$i\gets 1$ to $N$}\Comment{Iterate $N$ times}
    \State{$powers.append(pm.get\_power())$}
    \EndFor
    \State{$out.write(att\_list.get\_att(), powers, init\_rng)$}
    \Comment{Write to file}
    \State{$att.set\_att(attval)$}\Comment{Set attenuator $att$ to $attval$}
    \State{$pm.set\_range(rngval)$}\Comment{Change range of MPM}
    \State{$att\_list.disable()$}\Comment{Disable all attenuators}
    \State{$pm.zero()$}\Comment{Zero the MPM}
    \State{$att\_list.enable()$}\Comment{Enable all attenuators}
    \State{$powers\gets \{\}$}
    \For{$i\gets 1$ to $N$}\Comment{Iterate $N$ times}
    \State{$powers.append(pm.get\_power())$}
    \EndFor
    \State{$out.write(att\_list.get\_att(), powers, rngval)$}
    \Comment{Write to file}
    \State{$att.set\_att(0)$}
    \EndFor
    \EndProcedure
  \end{algorithmic}
\end{algorithm}

\subsection{System detection efficiency estimation}

SDE measurements are done with the detector port spliced to the fiber
coupled to the DUT. Since attenuator calibrations are integrated with this
routine, the optical switch will be used to route the light to either the
DUT or the MPM in different steps. The procedure is shown in algorithm
\ref{alg:sde}. We record counts from the DUT while scanning the voltage
bias applied. This is done under (a) ``dark'' circumstances to record the
dark counts, and under ``bright'' circumstances (laser-light directed at
DUT) (b) with polarization settings set to maximize the counts, and (c)
with the polarization settings set to minimize the counts. The SDE
measurements are only done with the \texttt{MCP-01} all-fiber polarization
controller, whose transmission loss is independent of polarization
settings. The optical switch calibration, the MPM nonlinearity correction
calibration (see algorithm \ref{alg:nonlin}), and the SDE counts
measurement (see algorithm \ref{alg:sde}) for any wavelength can be
performed in any order, but all three are essential for SDE estimation.

For an estimated absolute detector-port power of $P_{\text{DP}}$ (with all
three attenuators set to 'zero-attenuation' setting), and with the
attenuations calibrated to be $(\alpha_1, \alpha_2, \alpha_3)$, the number of photons
exiting the detector port per unit time with the attenuations applied would
be

\begin{equation}
  N_{\text{photons}}=\frac{P_{\text{DP}}\cdot\alpha_1\cdot\alpha_2\cdot\alpha_3\times\lambda}{hc},
\end{equation}

where $hc/\lambda$ is the energy of a single photon at wavelength $\lambda$. The SDE at
any given current bias is estimated by dividing the net count rate of the
DUT by $N_{\text{photons}}$. The net count rate is defined as the
difference between the time-averaged count rates ($\langle CR\rangle$) with the light
routed to the DUT and the time-averaged dark-count rates ($\langle DCR\rangle$). The
relative uncertainty in SDE is expressible as

\begin{equation}
  \frac{\sigma_{\text{SDE}}}{SDE}=\sqrt{\frac{\sigma_{\text{CR}}^2+\sigma_{\text{DCR}}^2}{(\langle
      CR\rangle-\langle DCR\rangle)^2}+\left(\frac{\sigma_{\text{DP}}}{P_{\text{DP}}}\right)^2+\sum\limits_{i=1}^3\left(\frac{\sigma_{\alpha_i}}{\alpha_i}\right)^2},
\end{equation}

where the relative uncertainty in the estimated detector-port power is in
turn

\begin{equation}
  \frac{\sigma_{\text{DP}}}{P_{\text{DP}}}=\sqrt{\left(\frac{\sigma_{\text{MPM}}}{P_{\text{MPM}}}\right)^2+\left(\frac{\sigma_{\text{SW}}}{R_{\text{SW}}}\right)^2+\left(\frac{\sigma_{\text{CPM}}}{CF_{\text{CPM}}}\right)^2+\left(\frac{\sigma_{\text{NL}}}{CF_{\text{NL}}}\right)^2}.
\end{equation}

\begin{algorithm}[h!]
  \caption{SDE counts measurement}\label{alg:sde}
  \begin{algorithmic}[1]
    \Procedure{console\_DE}{$device\_list,params,out$}
    \Comment{$params\equiv$ parameters, $out\equiv$ files}
    \State{$att\_list\gets device\_list[0,1,2]$}\Comment{List of attenuators}
    \State{$[pm, sw, pc]\gets device\_list[3,4,5]$}
    \Comment{$pm\equiv$ MPM, $sw\equiv$ optical switch, $pc\equiv$ polarization controller}
    \State{$[vsrc, counter]\gets device\_list[6,7]$}
    \Comment{$vsrc\equiv$ voltage source, $counter\equiv$ pulse counter}
    \State{}
    \State{$[attval, rngval]\gets params[0,1]$}\Comment{$attval\equiv$ attenuator
      value, $rngval\equiv$ MPM range value}
    \State{$[vstop, vstep, vpol]\gets params[2,3,4]$}\Comment{Voltage bias
      end-stop, step, and polarization optimization settings}
    \State{$[out\_att, out\_maxpol, out\_minpol]\gets out$}
    \Comment{Output files}
    \State{}
    \State{$N=10$}\Comment{Number of counter readings}
    \State{$att\_list.set\_att(attval)$}\Comment{Set attenuators to $attval$}
    \State{$vsrc.set\_volt(0)$}\Comment{Set voltage bias to $0$ V}
    \State{}
    \State{$[out\_maxpol, out\_minpol].write($'\# Dark Counts'$)$}
    \Comment{Start recording dark counts}
    \State{$sw.set\_route($'monitor\_port'$)$}
    \State{$att\_list.disable()$}
    \For{$vval\gets 0$ to $vstop$ by $vstep$}
    \State{$vsrc.set\_volt(vval)$}
    \For{$i\gets 1$ to $N$}\Comment{Iterate $N$ times}
    \State{$counts = counter.get\_counts()$}
    \State{$[out\_maxpol, out\_minpol].write(vval, counts)$}
    \Comment{Count for 1 sec., write to files}
    \EndFor
    \EndFor
    \State{$vsrc.set\_volt(0)$}
    \State{}
    \State{$att\_list.enable()$}
    \State{$sw.set\_route($'detector\_port'$)$}
    \State{$vsrc.set\_volt(vpol)$}\Comment{Set voltage bias for
      polarization optimization}
    \State{$maxpol\_settings\gets maximize\_counts(pc,counter,out\_maxpol)$}
    \State{$minpol\_settings\gets minimize\_counts(pc,counter,out\_minpol)$}
    \State{$vsrc.set\_volt(0)$}
    \State{}
    \State{$out\_maxpol.write($'\# Maxpol light counts'$)$}
    \Comment{Start recording 'maxpol' counts}
    \State{$pc.set(maxpol\_settings)$}
    \For{$vval\gets 0$ to $vstop$ by $vstep$}
    \State{$vsrc.set\_volt(vval)$}
    \For{$i\gets 1$ to $N$}\Comment{Iterate $N$ times}
    \State{$counts = counter.get\_counts()$}
    \State{$out\_maxpol.write(vval, counts)$}
    \Comment{Count for 1 sec., write to files}
    \EndFor
    \EndFor
    \State{$vsrc.set\_volt(0)$}
    \State{}
    \State{$out\_minpol.write($'\# Minpol light counts'$)$}
    \Comment{Start recording 'minpol' counts}
    \State{$pc.set(minpol\_settings)$}
    \For{$vval\gets 0$ to $vstop$ by $vstep$}
    \State{$vsrc.set\_volt(vval)$}
    \For{$i\gets 1$ to $N$}\Comment{Iterate $N$ times}
    \State{$counts = counter.get\_counts()$}
    \State{$out\_minpol.write(vval, counts)$}
    \Comment{Count for 1 sec., write to files}
    \EndFor
    \EndFor
    \State{$vsrc.set\_volt(0)$}
    \State{}
    \State{$sw.set\_route($'monitor\_port'$)$}
    \State{CONSOLE\_CAL$(pm,att\_list,attval,rngval,out\_att)$}
    \Comment{Attenuator calibration (see algorithm \ref{alg:att})}
    \EndProcedure
  \end{algorithmic}
\end{algorithm}

The standard deviations for the count rate ($\sigma_{\text{CR}}$) and the
dark-count rate ($\sigma_{\text{DCR}}$) are merely the Poisson standard
deviations ($\sqrt{\langle CR\rangle}$ and $\sqrt{\langle DCR\rangle}$ respectively). The numerical
standard deviations of the ten consecutive measurements (see algorithm
\ref{alg:sde}) made at every current bias was found to be slightly less
than the expected Poisson value. We will be using the quadrature sum of the
numerical standard deviation and the Poisson standard deviation in our
error analysis. Table \ref{tbl:err} tabulates the relative uncertainties
estimated for the other variables.

\begin{table}[h!]
  \caption{Error budget} \label{tbl:err}
  \centering
  \begin{tabular}{lcr}
    \toprule
    Source & Symbol & Standard ($k=1$) relative uncertainty\\
    \midrule
    Calibration factor & $CF_{\text{CPM}}(\lambda)$ & $0.14\%$\\
    Nonlinearity correction & $CF_{\text{NL}}(\lambda, r=-30\text{ dBm})$ & $0.075\%$\\
    Optical switching ratio & $R_{\text{SW}}$ & $0.14\%$\\
    Optical attenuation & $\alpha_i$, $i\in\{1, 2, 3\}$ & $0.2\%$\\
    Monitor power meter & $P_{\text{}MPM}\vert_{r=-30\text{ dBm}}$ & $0.1\%$\\
    \midrule
    SDE @ $2\times10^5$ counts per second && $0.46\%$\\
    SDE @ $1\times10^5$ counts per second && $0.51\%$\\
    \bottomrule
  \end{tabular}
\end{table}

Many of the source terms in table \ref{tbl:err} are composites of other
source terms, and consequently inherit their errors. Using these
conservative errors, the $k=1$ relative uncertainty for our SDE estimates
comes to $0.46\%$ at count rates of $2\times10^5$ per second, and $0.51\%$ at
count rates of $1\times10^5$ per second. Note that in the data presented in the
main manuscript, all the errors for SDE were programmatically computed
using the \verb|uncertainties| \verb|python| library, which automatically
tracks the error propagation (with covariances for any calculated fit
parameters) through arithmetic operations.

\begin{figure}[h!]
\centering
\includegraphics[width=\linewidth]{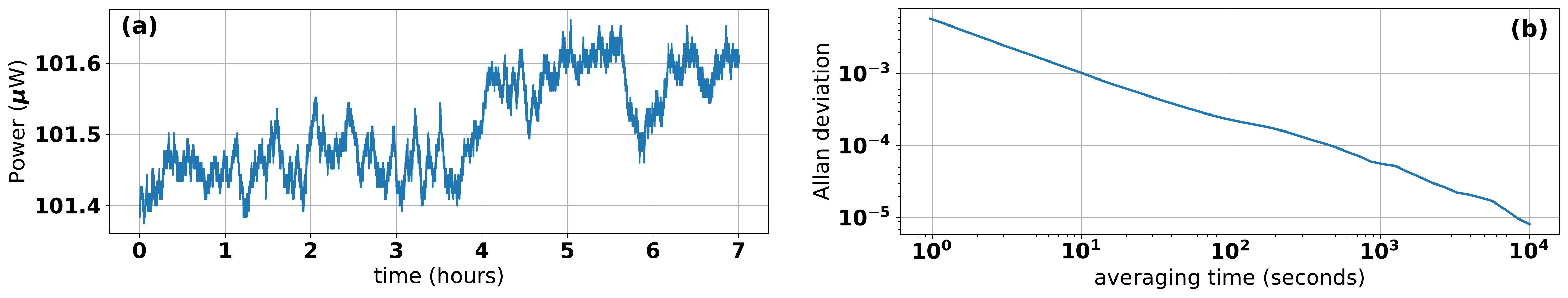}
\caption{\label{fig_supp8} (a) The power meter reading after $17$ hours of
  equilibration time, sampled at approximately $4.118$ Hz. (b) Allan
  deviation plot on a log-log scale for the same. Allan deviation for
  10-second averaging is $= 9.28e-4$. }
\end{figure}

All the optical components used in SDE estimation were measured to be
stable over a period of 10 hours. The laser had a built-in power setting
and a wavelength setting. We set the power setting and let it equilibrate
for over 24 hours before beginning any measurements. Changing the
wavelength changed the power, but did not introduce any additional drift.
As a precaution the wavelength changes were also dwelled on for half an
hour of equilibration. Figure \ref{fig_supp8}(a) shows the laser power
drift (on the MPM) after a power setting was set and the laser was allowed
to equilibrate for 17 hours. The typical duration for a complete wavelength
scan per DUT was 7 hours, a large portion of which was meant for
equilibration dwells for wavelength setting changes. Figure
\ref{fig_supp8}(b) shows the Allan deviation for the laser power
\cite{Witt2001}. The deviation for averaging timescales of 10 seconds was
below 0.1\%.

\section{Polarization sensitivity measurement}

In order to ensure that we were sampling all input polarization states on
the Bloch sphere, we replaced the \texttt{MCP-01} all-fiber polarization
controller with an \texttt{HP8169A} free-space polarization controller.
This instrument out-coupled the fiber input to a free-space beam that is
transmitted successively through a linear polarizer, a quarter-wave plate,
and a half-wave plate, and then coupled back into an output fiber port. The
three optical components are mounted on motorized rotational mounts,
allowing for complete control of the polarization state of the output mode.
Unlike the all-fiber polarization controller, this instrument had a
polarization-setting dependent transmission loss, which had to be measured
using both the CPM and MPM by splicing the detector-port to the fiber
coupled to the CPM. The free-space nature of the instrument introduced
significant drift on an hour timescale. Therefore this
polarization-settings-dependent transmission was estimated for each of the
three DUT counts-based measurements in turn.

\begin{figure}[h!]
\centering
\includegraphics[width=\linewidth]{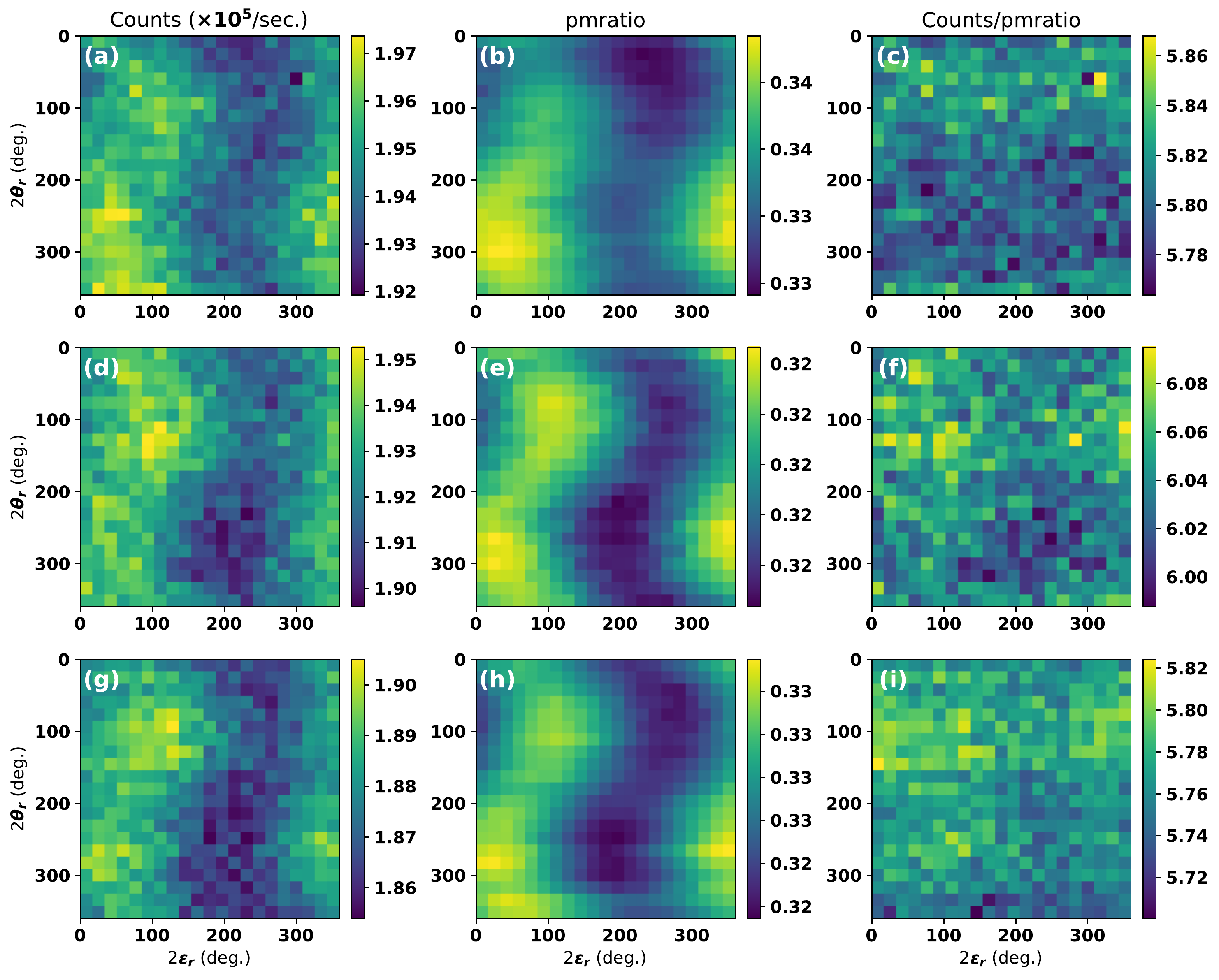}
\caption{\label{fig_supp9} The raw net counts (light counts - dark counts)
  (a, d, g), the power-meter ratio or transmission factor (b, e, h), and
  the transmission-factor corrected count rate (c, f, i) for detectors D1
  (a, b, c), D2 (d, e, f), and D3 (g, h, i) over various polarization
  settings on the \texttt{HP8169A} free-space polarization controller. The
  max/min polarization sensitivities are (c) 1.018, (f) 1.018, and (i)
  1.021. The standard uncertainty for all three is $\pm 0.008$.}
\end{figure}

Figure \ref{fig_supp9} plots the raw counts, the polarization-settings
dependent transmission loss (expressed as 'pmratio'), and the
loss-normalized counts for all three DUTs in three separate rows. No
attempt at smoothing the data was made when estimating polarization
sensitivity in this method.

\section*{Data and code availability statement}

All of the data gathered for this manuscript and the \texttt{python} code
used to implement the processing described in this document has been made
available as a \texttt{zip}-archive file (\texttt{SHA-1
  91599a67964d30b695b50b4719985533e98869a0}), along with a digital
signature from the primary author (\texttt{gpg} fingerprint: \texttt{CC49
  3CCF 1104 0DC7 36C4 5A7C E4E5 0022 5ED1 7577}). These files are available
as supplementary materials through the journal, and copies of the same can
be provided by the authors upon request.

\section*{Disclaimer}

Identification of commercial instruments in this document does not imply
recommendation or endorsement by the National Institute of Standards and
Technology.

